\documentclass[a4,12pt]{article}%
\usepackage{amsfonts}
\usepackage{amsmath}
\usepackage{amssymb}
\usepackage{graphicx}%
\begin{document}

\title{Effective Field Equations of Brane-Induced Electromagnetism}
\author{F. Dahia$^{a}$, C. Romero$^{b}$ and M. A. S. Cruz$^{b}$,\\$^{a}$Departamento de F\'{\i}sica, Universidade Federal de Campina Grande, \\58109-970, Campina Grande, Pb, Brazil\\$^{b}$Departamento de F\'{\i}sica, Universidade Federal da Para\'{\i}ba,\\Caixa \ Postal 5008, 58050-970, Jo\~{a}o Pessoa, Pb, Brazil}
\maketitle

\begin{abstract}
Using a covariant embedding formalism, we find the effective field equations
for the Electromagnetism that emerge on branes in the context of
Dvali-Gabadadze-Porrati (DGP) braneworld scenario. Our treatment is
essentially geometrical. We start with Maxwell equations in five-dimensions
and project them into an arbitrary brane. The formalism is quite general and
allows us to consider curved bulk spaces and curved branes whose tension is
not necessarily null. The kinetic electromagnetic term induced on the world
volume of the brane, proper of DGP models, is incorporated in this formulation
by means of an appropriate match condition. We also give an estimate of each
term of the effective field equations and determine the domain in which the
four-dimensional Maxwell equations can be recovered in the brane.

\end{abstract}

\section{Introduction}

According to recent braneworld scenarios all fields of the
Standard Model are trapped in a four-dimensional (4D) spacetime,
which is seen as a manifold embedded in a higher-dimensional space
\cite{ADD,IADD,Randall,RSII}. In this picture the ambient space,
or \textit{bulk}, appears to be more fundamental, in a sense, than
the ordinary four-dimensional spacetime. Therefore it seems
plausible that the basic laws of physics should primarily be
formulated in this higher-dimensional setting. According to this
view, the apparent validity of our familiar four-dimensional laws
would merely appear as a consequence of more fundamental
higher-dimensional laws \cite{Shiromizu,maeda,Maia,gonta}. This is
very much in the spirit of the original Kaluza-Klein theory and
its modern versions where the fundamental field equations, which
unify gravity and electromagnetism, are formulated in a
higher-dimensional manifold \cite{KK}. Another proposal that
essentially\ follows the same principle is known as the
induced-matter approach \cite{Leon,Overduim,Wesson}. Here the
fundamental field equations are again five-dimensional (5D) and\
four-dimensional macroscopic matter configurations are regarded as
a purely geometrical manifestation of the extra dimension. Of
course the "embedding" \ of\ a four-dimensional theory into five
or higher dimensions is naturally subjected to the so-called
embedding theorems \cite{Goenner}. In the case of the
induced-matter approach the embedding is guaranteed by the
well-known Campbell-Magaard theorem
\cite{campbell,magaard,romero,dahia1,dahia2,dahia2,dyn}, whereas
in the case of branes an extended version of the latter is the
relevant mathematical result to be used \cite{seahra,Dahia}.

A fundamental ingredient in the brane world scenarios is the mechanism of
field localization. In the context of string theories these mechanisms arises
very naturally \cite{polchinsky}. But, there also exist methods of
localization based on field-theoretic framework such as the capture of
fermions by topological defects \cite{rubakov,akama}. In this paper we are
going to consider localization mechanisms of vector fields provided by
field-theoretic means. However, in a sense, the confinement of massless gauge
fields requires a more intricate mechanism than those methods applicable to
matter fields. The reason is that gauge fields trapped by topological defects
will not be massless unless new elements be added to the theory \cite{shifman}%
. Furthermore, as it was pointed out in Ref. \cite{rubakov2}, the confinement
of gauge fields should preserve the universality of charge in the brane, but a
hypothetical mechanism that was based on a direct localization of the zero
mode in topological defects would have little chance to fulfill this principle.

There are alternative approaches which avoid these problems satisfactorily.
Among them, we can mention models in which the confinement is performed by
gravity \cite{oda1,rubakov3}. As a matter of fact, if the bulk is a
five-dimensional warped space, it is known that gravity is not strong enough
to keep the vector field localized in the brane \cite{pomarol,bajc}. However,
the confinement can work appropriately by adding new warped extra dimensions
which are also compact. Other models follow a complete different approach.
They propose a modification of the gauge field equations in the bulk in order
to achieve the localization in the brane
\cite{shifman,oda2,evgeny,neronov,dick}.

In this paper we are going to focus our discussion on the well-known DGP model
\cite{DGP}, because of its generality (it can be employed in any theories with
extra-dimensions) and interesting physical implications, specially, for
cosmology \cite{cosmos} (such as the possibility of explaining the late
accelerated phase of expansion of the Universe). The appearance of DGP model
might be seen as a further step in the road of extra dimension theories.
Indeed, roughly speaking, we might say that the history of these theories
begins with Kaluza-Klein's idea of tiny compact dimensions and develops
towards the RSII model of infinite large dimension \cite{RSII}. It turns out
that within the RSII model the proper world volume integrated along the extra
dimension is finite because of the warping factor. With DGP model this last
constraint is abandoned and we are led to a consistent large extra dimension
model which is also world volume infinite \cite{DGP}.

The basic assumption of DGP model is that radiative quantum corrections
associated to the interactions between the bulk gauge fields and localized
matter in the brane might be effectively incorporated in the theory by adding
a new term in the Lagrangean, which corresponds to a kinetic term for the bulk
fields, restricted to the world volume of the brane.

Our main goal in this paper is to obtain a covariant formulation of the
effective equation for the electromagnetism in DGP scenario. In other words,
we are interested in finding out how 4D-observers describe the behavior of the
electromagnetic field, considering that they live confined in a brane (with a
not necessarily null tension) in the context of DGP scenario. We will start
with a five-dimensional version of Maxwell equations, which we will assume to
hold in the bulk, and by using the formalism of embedding theory we will
determine the induced equation in the brane. The covariant formalism used here
is very general and embodies a great variety of geometric configurations.
Indeed, except for the reflection symmetry, there is no restriction imposed on
the geometry of the bulk and brane's manifolds.

As we shall see, the effective field equations contains additional terms as
compared to the usual Maxwell equations and possess two important parameters:
the brane tension $\left(  \lambda\right)  $ and the five-dimensional
electromagnetic coupling constant. We are going to suggest a possible physical
interpretation for the new terms from the point of view of the 4D-observers
and give an estimate for them relatively to the parameters.

By considering the electromagnetic coupling constant, the ratio between the
four-dimensional and the five-dimensional values establishes a length scale in
the DGP model $\left(  r_{c}\right)  $ \cite{DGP2}. A very peculiar feature of
DGP model is the prediction that deviations from the four-dimensional behavior
occurs at ultra-large distances as compared to $r_{c}$, unlike other extra
dimension theories.

In a brane with a non-zero tension, there will be two length scales regulating
the effective equations: $\left(  G_{5}\lambda\right)  ^{-1}$ and $r_{c}$,
where $G_{5}$ is the five-dimensional gravitational constant. Of course, the
induced equations on the brane should be consistent with all experimental data
related to electromagnetic phenomena. Based on our estimates we are going to
study the dependence of the effective field equations on these parameters and
check under what condition Maxwell equations could be recovered.

The paper is organized as follows. In Section 2 we present the basics of the
embedding formalism and the techniques of projecting general tensors on
hypersurfaces. In Section 3, we establish a relation between the intrinsic and
the projected electromagnetic field in a hypersurface, by means of which we
find the effective equations in the hypersurface. We proceed with Section 4,
where we discuss the question whether any solution of the 4D Maxwell equations
might be reproduced from five dimensions. In section 5, assuming the
reflection symmetry and using the appropriate match condition we obtain the
effective field equations for the electromagnetic field in DGP scenario. Here
we also discuss the low-energy limit of the effective equations. Finally,
Section 5 \ contains our final remarks.

\section{The embedding formalism}

In this section we shall employ some techniques from differential geometry in
order to set the basic equations that regulates the embedding of the
four-dimensional electromagnetic equations in higher dimensions. We start by
recalling some definitions.

Let us assume that our $n$-dimensional spacetime $\Sigma$ is embedded into a
space $\hat{M}$ with $\left(  n+1\right)  $ dimensions and let $\phi
:\Sigma\rightarrow\hat{M}$ be the embedding map. Then, by definition of
embedding, the image $\phi\left(  \Sigma\right)  $ is a hypersurface of
$\hat{M}$ \cite{Manfredo}. Now, if $\Sigma$ and $\hat{M}$ are equipped with
the metrics $g$ and $\tilde{g}$, respectively, then the embedding is said to
be isometric if the intrinsic metric of $\ \Sigma$ corresponds to the induced
metric in the hypersurface, i.e., if the condition%
\begin{equation}
g\left(  \mathbf{v},\mathbf{w}\right)  =\hat{g}\left(  d\phi\left(
\mathbf{v}\right)  ,d\phi\left(  \mathbf{w}\right)  \right)  \label{isometric}%
\end{equation}
holds for any vectors $\mathbf{v}$ and $\mathbf{w}$ of the tangent \ space
$T_{p}\Sigma$ for every $p\in\Sigma$ \footnote{Here we are adopting the
following notation: a small bold letter, as for example $\mathbf{v}$,
represents a vector of the tangent space of $M$ while a capital bold letter
$\left(  \mathbf{V}\right)  $ corresponds to a vector of the tangent space of
$\hat{M}$.}. Since $\phi$ is a homeomorphism let us, henceforth, identify
$\Sigma$ with $\phi\left(  \Sigma\right)  $ and as the differential map
$d\phi$ is injective there is no confusion if we write simply $\mathbf{v}$ in
place of $d\phi\left(  \mathbf{v}\right)  .$ Now $\ $let $\left\{
y^{A}\right\}  $ and $\left\{  x^{\mu}\right\}  $ be local coordinate systems
of $\hat{M}$ and $\Sigma,$ respectively \footnote{Throughout capital Latin
indices take value in the range (0,1,...,n) while Greek indices run from
(0,1,...,n-1).}. The embedding map in terms of these coordinates is described
by the functions $y^{A}=\phi^{A}\left(  x\right)  $. Of course, the image of
any vector $\mathbf{v\in}T_{p}\Sigma$ belongs also to the tangent space
$T_{p}\hat{M}$ of the ambient space $\hat{M}$ and, then, can be written in
terms of the coordinate basis $\left\{  \partial_{A}\right\}  $\footnote{Here
we are using the common notation $\partial_{A}=\frac{\partial}{\partial y^{A}%
}$ and $\partial_{\alpha}=\frac{\partial}{\partial x^{\alpha}}$. We shall
represent the components of vectors with respect to these bases as $v^{\alpha
}$ and $V^{A}$, respectively.}. In particular, we can write $\partial_{\alpha
}=e_{\alpha}^{A}\partial_{A}$, where%

\begin{equation}
e_{\alpha}^{A}=\frac{\partial\phi^{A}}{\partial x^{\alpha}} \label{dphi}%
\end{equation}
are the elements of the differential map $d\phi$ written with respect to the
given coordinates. The isometric condition (\ref{isometric}) expressed
relatively to the coordinate bases takes the form%
\begin{equation}
g_{\alpha\beta}=e_{\alpha}^{A}e_{\beta}^{B}\hat{g}_{AB} \label{alfabeta}%
\end{equation}

Let us now consider the normal vector of the hypersurface $\Sigma$ with
respect to $\hat{M}.$It can be obtained, up to orietation, by solving the
following equations:%
\begin{align}
\hat{g}\left(  \partial_{\alpha},\mathbf{N}\right)   &  =0\Rightarrow
e_{\alpha}^{A}N_{A}=0\\
\hat{g}\left(  \mathbf{N},\mathbf{N}\right)   &  =\varepsilon=+1
\end{align}
where we are admitting that the normal vector is spacelike ($\varepsilon=+1$,
according to our convention). It follows from the injective condition of the
differential map that the rank of $e_{\alpha}^{A}$ is equal to the dimension
of $\Sigma$, and as a consequence there is only one (the number of the
codimension) linearly independent normal vector satisfying the above
equations. Let us denote it by $\mathbf{N}$.

In the present formalism an important concept is that of a projection tensor
$\Pi$, which maps vectors $\mathbf{V}\in T_{p}(\hat{M})$ onto the tangent
space $T_{p}(\Sigma)$ of the hypersurface at $p\in$ $\Sigma.$ We define $\Pi$
by%
\begin{equation}
\Pi\left(  \mathbf{V}\right)  =\mathbf{V}-V^{\perp}\mathbf{N} \label{pi}%
\end{equation}
where here we are using the following notation $V^{\perp}=\hat{g}\left(
\mathbf{V},\mathbf{N}\right)  $.

It is clear that $\Pi\left(  \mathbf{V}\right)  $ can be considered as a
vector of $T_{p}\Sigma$. Therefore it can be written in the basis $\left\{
\partial_{\alpha}\right\}  $. In particular, the projection of $\partial
_{A}\in T_{p}\hat{M}$ can be written as $\Pi\left(  \partial_{A}\right)
=e_{A}^{\alpha}\partial_{\alpha}$, for some 'vielbein' $e_{A}^{\alpha}$. As we
shall see later, $e_{A}^{\alpha}$ and $e_{\alpha}^{A}$ are related. Using
(\ref{pi}) we can write vector $\partial_{A}$ as%
\begin{equation}
\partial_{A}=e_{A}^{\alpha}\partial_{\alpha}+N_{A}\mathbf{N} \label{DA}%
\end{equation}
Note that this relation is valid only at points of $\Sigma$, where $\left\{
\partial_{\alpha},\mathbf{N}\right\}  $ constitute a local basis of
$T_{p}\left(  \hat{M}\right)  $.

To determine the relation between the vielbein functions $e_{A}^{\alpha}$ and
$e_{\alpha}^{A}$ let us consider the vectors $\partial_{\beta}$ and
$\partial_{A}$, and calculate their inner product $\hat{g}\left(
\partial_{\beta},\partial_{A}\right)  $. From (\ref{DA})\ it is easy to see
that $\hat{g}\left(  e_{\beta}^{B}\partial_{B},\partial_{A}\right)  =\hat
{g}\left(  \partial_{\beta},e_{A}^{\alpha}\partial_{\alpha}\right)  $, which
then shows that \
\begin{equation}
e_{A}^{\alpha}=\hat{g}_{AB}g^{\alpha\beta}e_{\beta}^{B}%
\end{equation}

Now let us turn our attention to the decomposition of an arbitrary vector
$\mathbf{V}\in T_{p}\hat{M}$. We have $\mathbf{V}=V^{A}$ $\partial_{A}$ and
from (\ref{DA}) it may be written as a sum of parallel and orthogonal
components with respect to the hypersurface $\Sigma$:
\begin{equation}
\mathbf{V}=V^{\alpha}\partial_{\alpha}+V^{\perp}\mathbf{N} \label{V}%
\end{equation}
where $V^{\alpha}=e_{A}^{\alpha}V^{A}$ corresponds to the coordinates of the
projected vector, $\Pi\left(  \mathbf{V}\right)  $, in the basis $\left\{
\partial_{\alpha}\right\}  $. It is also useful to express explicitly the
relation between the components of the vector in the bases $\left\{
\partial_{A}\right\}  $ and $\left\{  \partial_{\alpha},\mathbf{N}\right\}  $:%
\begin{equation}
V^{A}=e_{\alpha}^{A}V^{\alpha}+V^{\perp}N^{A} \label{VA}%
\end{equation}

Now, with the help of the projection operator $\Pi$, tensors fields of any
rank can be decomposed with respect to the basis $\left\{  \partial_{\alpha
},\mathbf{N}\right\}  $ at points of $\Sigma$. In particular, using
(\ref{DA}), the second rank tensor $\mathbf{T}=T^{AB}\partial_{A}%
\otimes\partial_{B}$ can be written as%

\begin{equation}
\mathbf{T}=T^{AB}\partial_{A}\otimes\partial_{B}=T^{\alpha\beta}%
\partial_{\alpha}\otimes\partial_{\beta}+\psi^{\alpha}\partial_{\alpha}%
\otimes\mathbf{N}+\chi^{\beta}\mathbf{N}\otimes\partial_{\beta}+\phi
\mathbf{N}\otimes\mathbf{N}%
\end{equation}
where
\begin{align}
T^{\alpha\beta}  &  =e_{A}^{\alpha}e_{B}^{\beta}T^{AB}\\
\psi^{\alpha}  &  =e_{A}^{\alpha}N_{B}T^{AB}\label{psi}\\
\chi^{\beta}  &  =e_{B}^{\beta}N_{A}T^{AB}\label{qi}\\
\phi &  =N_{A}N_{B}T^{AB} \label{phi}%
\end{align}
Notice that the components $T^{\alpha\beta}$ are just the components of the
projection of $\mathbf{T}$ with respect to the intrinsic coordinates basis
$\{\partial_{\alpha}\}$\ of the tangent space of the hypersurface $\Sigma$.
Clearly, the components of $\mathbf{T}$ $\ $in both coordinate bases\ are
related by the equation
\begin{equation}
T^{AB}=e_{\alpha}^{A}e_{\beta}^{B}T^{\alpha\beta}+e_{\alpha}^{A}N^{B}%
\psi^{\alpha}+N^{A}e_{\beta}^{B}\chi^{\beta}+N^{A}N^{B}\phi
\end{equation}

We now turn to the question of how differential operators acting on $\Sigma
\ $\ and $\hat{M}$\ \ are related. Let $D$ and $\nabla$ denote the covariant
derivative relative to the metrics $\hat{g}$ and $g$ respectively. In the case
of an isometric embedding $\nabla$ corresponds to the induced covariant
derivative \cite{Manfredo}:%
\begin{equation}
\nabla_{\mathbf{w}}\mathbf{v}=\Pi\left(  D_{\mathbf{w}}\mathbf{v}\right)
\label{CovD}%
\end{equation}
where now $\mathbf{v}$ and $\mathbf{w}$ are two vectors fields of the tangent
space of $\Sigma$. (It is important to note that in the expression
$D_{\mathbf{w}}\mathbf{v}$, instead of $\mathbf{v}$ and $\mathbf{w}$, we
should consider any extensions of these vector fields \cite{Manfredo}).

Now take $\mathbf{w}=\partial_{\beta}$ and write $\mathbf{v}$ in the basis
$\left\{  \partial_{A}\right\}  $. From (\ref{CovD}) we obtain%
\begin{equation}
\nabla_{\beta}v^{\alpha}=e_{A}^{\alpha}e_{\beta}^{B}D_{B}v^{A}%
\end{equation}

Another important concept in the context of the embedding formalism is that of
extrinsic curvature $K$, which can be defined from the orthogonal components
of the vector $D_{\mathbf{w}}\mathbf{v}$ as%
\begin{equation}
K\left(  \mathbf{v},\mathbf{w}\right)  \equiv\hat{g}\left(  D_{\mathbf{w}%
}\mathbf{v},\mathbf{N}\right)  =-\hat{g}\left(  \mathbf{v},D_{\mathbf{w}%
}\mathbf{N}\right)  \label{K}%
\end{equation}
(The latter equality comes from the condition of compatibility between
$\hat{g}$ and the covariant derivative $D$). It is easy to see that the
extrinsic curvature is a symmetric tensor whose components in the coordinates
$\left\{  \partial_{\alpha}\right\}  $ can be written as
\begin{equation}
K_{\alpha\beta}=-e_{\alpha}^{A}e_{\beta}^{B}D_{B}N_{A}%
\end{equation}

Now let us consider the covariant derivative of the normal vectors
$\mathbf{N}$ along tangent directions of $\Sigma$. Relatively to the basis
$\left\{  \partial_{\alpha},\mathbf{N}\right\}  $, we find the following
decomposition:%
\begin{equation}
D_{\beta}\mathbf{N}=-K_{\beta}^{\alpha}\partial_{\alpha} \label{DN}%
\end{equation}
Now we have all the necessary ingredients to analyze the decomposition of the
covariant derivative $D_{B}\mathbf{V}$ of an arbitrary vector field
$\mathbf{V\ }$of the tangent space $\hat{M}$. Using (\ref{DA}), we can split
the covariant derivative in two terms: one related to derivative along the
tangent directions of $\Sigma$ and the other one, related to the normal
direction:
\begin{equation}
D_{B}\mathbf{V}=e_{B}^{\beta}D_{\beta}\mathbf{V}+N_{B}D_{\perp}\mathbf{V}
\label{DV}%
\end{equation}
where we have introduced the following notation: $D_{\perp}\mathbf{V}%
=D_{\mathbf{N}}\mathbf{V}$. As the vector $\mathbf{V}$ itself can be also
decomposed according to (\ref{V}), then, the first term on the right hand side
of (\ref{DV}) gives%
\begin{equation}
D_{\beta}\mathbf{V}=D_{\beta}\left(  V^{\alpha}\partial_{\alpha}\right)
+D_{\beta}\left(  V^{\perp}\mathbf{N}\right)
\end{equation}
Using (\ref{CovD}), (\ref{K}) and (\ref{DN}), it follows that%
\begin{align}
D_{\beta}\left(  V^{\alpha}\partial_{\alpha}\right)   &  =\nabla_{\beta
}\left(  V^{\alpha}\partial_{\alpha}\right)  +K_{\alpha\beta}V^{\alpha
}\mathbf{N}\\
D_{\beta}\left(  V^{\perp}\mathbf{N}\right)   &  =\mathbf{N}\nabla_{\beta
}V^{\perp}-V^{\perp}K_{\beta}^{\alpha}\partial_{\alpha}%
\end{align}
Therefore, the derivative of the vector $\mathbf{V}$ along the tangent
direction $\Pi\left(  \partial_{B}\right)  $ has the following components in
the coordinates $\left\{  \partial_{A}\right\}  $:%
\begin{align}
e_{B}^{\beta}D_{\beta}V^{A}  &  =e_{B}^{\beta}\left\{  e_{\alpha}^{A}\left[
\nabla_{\beta}V^{\alpha}-K_{\beta}^{\alpha}V^{\perp}\right]  \right.
\nonumber\\
&  \left.  +N^{A}\left[  \nabla_{\beta}V^{\perp}+K_{\alpha\beta}V^{\alpha
}\right]  \right\}
\end{align}

As we have already mentioned, the second term in (\ref{DV}) corresponds to the
covariant derivative relative to the normal direction. Thus, collecting all
terms, we find%
\begin{align}
D_{B}V^{A}  &  =e_{B}^{\beta}e_{\alpha}^{A}\left[  \nabla_{\beta}V^{\alpha
}-V^{\perp}K_{\beta}^{\alpha}\right] \nonumber\\
&  +e_{B}^{\beta}N^{A}\left[  \nabla_{\beta}V^{\perp}+K_{\alpha\beta}%
V^{\alpha}\right]  +N_{B}D_{\perp}V^{A} \label{DBVA}%
\end{align}
\qquad\qquad

In similar manner we can obtain the decomposition of the covariant derivative
of the tensor $\mathbf{T}=T^{AB}\partial_{A}\otimes\partial_{B}$. Thus we have%
\begin{equation}
D_{C}\left(  T^{AB}\partial_{A}\otimes\partial_{B}\right)  =e_{C}^{\gamma
}D_{\gamma}\left(  T^{AB}\partial_{A}\otimes\partial_{B}\right)
+N_{C}D_{\perp}\left(  T^{AB}\partial_{A}\otimes\partial_{B}\right)
\label{DT}%
\end{equation}
Initially let us we consider the first term on the right-hand side of the
above equation:%
\begin{equation}
D_{\gamma}\left(  T^{AB}\partial_{A}\otimes\partial_{B}\right)  =D_{\gamma
}\left(  T^{\alpha\beta}\partial_{\alpha}\otimes\partial_{\beta}+\psi^{\alpha
}\partial_{\alpha}\otimes\mathbf{N}+\chi^{\beta}\mathbf{N}\otimes
\partial_{\beta}+\phi\mathbf{N}\otimes\mathbf{N}\right)
\end{equation}
Each term on the right-hand side of the latter equation can be analyzed
separately:%
\begin{align}
D_{\gamma}\left(  T^{\alpha\beta}\partial_{\alpha}\otimes\partial_{\beta
}\right)   &  =\nabla_{\gamma}\left(  T^{\alpha\beta}\partial_{\alpha}%
\otimes\partial_{\beta}\right)  +T^{\alpha\beta}K_{\alpha\gamma}%
\mathbf{N}\otimes\partial_{\beta}+T^{\alpha\beta}K_{\gamma\beta}%
\partial_{\alpha}\otimes\mathbf{N}\\
D_{\gamma}\left(  \psi^{\alpha}\partial_{\alpha}\otimes\mathbf{N}\right)   &
=\nabla_{\gamma}(\psi^{\alpha}\partial_{\alpha})\otimes\mathbf{N}+\psi
^{\alpha}K_{\alpha\gamma}\mathbf{N}\otimes\mathbf{N}\nonumber\\
&  -\psi^{\alpha}K_{\gamma}^{\beta}\partial_{\alpha}\otimes\partial_{\beta}\\
D_{\gamma}\left(  \chi^{\beta}\mathbf{N}\otimes\partial_{\beta}\right)   &
=\mathbf{N}\otimes\nabla_{\gamma}(\chi^{\beta}\partial_{\beta})+\chi^{\beta
}K_{\beta\gamma}\mathbf{N}\otimes\mathbf{N}\nonumber\\
&  -\chi^{\beta}K_{\gamma}^{\alpha}\partial_{\alpha}\otimes\partial_{\beta}\\
D_{\gamma}\left(  \phi\mathbf{N}\otimes\mathbf{N}\right)   &  =\left(
\nabla_{\gamma}\phi\right)  \mathbf{N}\otimes\mathbf{N}-\phi K_{\gamma
}^{\alpha}\partial_{\alpha}\otimes\mathbf{N}\nonumber\\
&  -\phi K_{\gamma}^{\beta}\mathbf{N}\otimes\partial_{\beta}%
\end{align}
Thus we have the following:%
\begin{align}
e_{C}^{\gamma}D_{\gamma}T^{AB}  &  =e_{C}^{\gamma}\left\{  e_{\alpha}%
^{A}e_{\beta}^{B}\left[  \nabla_{\gamma}T^{\alpha\beta}-\psi^{\alpha}%
K_{\gamma}^{\beta}-\chi^{\beta}K_{\gamma}^{\alpha}\right]  \right. \nonumber\\
&  +N^{A}e_{\beta}^{B}\left[  \varepsilon T^{\alpha\beta}K_{\alpha\gamma
}+\nabla_{\gamma}\chi^{\beta}-\phi K_{\gamma}^{\beta}\right] \nonumber\\
&  +e_{\alpha}^{A}N^{B}\left[  T^{\alpha\beta}K_{\gamma\beta}+\nabla_{\gamma
}\psi^{\alpha}-\phi K_{\gamma}^{\alpha}+\psi^{\alpha}\theta_{\gamma}%
^{cb}\right] \nonumber\\
&  +\left.  N^{A}N^{B}\left[  \psi^{\alpha}K_{\alpha\gamma}+\chi^{\beta
}K_{\beta\gamma}+\left(  \nabla_{\gamma}\phi\right)  \right]  \right\}
\label{DTAB}%
\end{align}
With the help of relation (\ref{DTAB}) and writing the components in second
term of (\ref{DT}) as $D_{\perp}T^{AB}$ we can establish a relation between
the components of the covariant derivative of the tensor $\mathbf{T}$ in the
bases $\left\{  \partial_{A}\right\}  $ and $\left\{  \partial_{\alpha
},\mathbf{N}\right\}  $: $D_{C}T^{AB}=e_{C}^{\gamma}D_{\gamma}T^{AB}%
+N^{A}D_{\perp}T^{AB}$, where $e_{C}^{\gamma}D_{\gamma}T^{AB}$ is given by
(\ref{DTAB}).

In the next section we are going to discuss the induction of the
higher-dimensional electromagnetic equations into the hypersurface $\Sigma$.
The higher-dimensional equations shall be prescribed as a higher-dimensional
generalization of the usual Maxwell equations. In this case, the field
equations will involve the divergence of an anti-symmetric second order tensor
which will play the role of the electromagnetic field. Then, it is important
to know how the divergence of a tensor is treated in the present embedding
formalism. So let us take the contraction between the indices $A$ and $C$ in
the covariant derivative of the tensor $\mathbf{T}$. In the local coordinates,
we have:%
\begin{equation}
D_{A}T^{AB}=e_{A}^{\gamma}D_{\gamma}T^{AB}+N_{A}D_{\perp}T^{AB}%
\end{equation}
Using the relation (\ref{DTAB}) and remembering that $e_{\alpha}^{A}%
e_{A}^{\gamma}=\delta_{\alpha}^{\gamma}$ and $e_{A}^{\gamma}N^{A}=0$, we find:%
\begin{align}
D_{A}T^{AB}  &  =e_{\beta}^{B}\left[  \nabla_{\alpha}T^{\alpha\beta}%
-\psi^{\alpha}K_{\alpha}^{\beta}-\chi^{\beta}K\right] \nonumber\\
&  +N^{B}\left[  T^{\alpha\beta}K_{\gamma\beta}+\nabla_{\alpha}\psi^{\alpha
}-\phi K\right]  +N_{A}D_{\perp}T^{AB} \label{divT}%
\end{align}
where $K=K_{\alpha}^{\alpha}$ is the trace of the extrinsic curvature.
Considering the parallel and orthogonal projection of the vector $D_{A}T^{AB}%
$\ with respect to $\Sigma$, we obtain, respectively:%
\begin{align}
\left(  D_{A}T^{AB}\right)  e_{B}^{\beta}  &  =\nabla_{\alpha}T^{\alpha\beta
}-\psi^{\alpha}K_{\alpha}^{\beta}-\chi^{\beta}K+W^{\beta}\label{divTtan}\\
\left(  D_{A}T^{AB}\right)  N_{B}  &  =T^{\alpha\beta}K_{\alpha\beta}%
+\nabla_{\alpha}\psi^{\alpha}-\phi K+Y \label{divTnormal}%
\end{align}
where $W^{\beta}=\left(  D_{\perp}T^{AB}\right)  N_{A}e_{B}^{\beta}$ and
$Y=\left(  D_{\perp}T^{AB}\right)  N_{A}N_{B}$. It is interesting to notice at
this point that $\chi^{\alpha},\psi^{\alpha}$ and $W^{\alpha}$ behave as
vectors with respect to the coordinates transformation of $\Sigma$, while
$\phi$ and $Y$ are scalar quantities.\

\section{The induced electromagnetic equations on the hypersurface}

In this section we shall investigate how a higher-dimensional electrodynamics
defined on the manifold $\hat{M}$ might induces four-dimensional
electromagnetism on the manifold $\Sigma$, (which we take to be the ordinary
space-time). So, let $\mathcal{A}^{A}$ be a higher-dimensional vector field,
which we assume to play the role of a higher-dimensional vector potential
defined on $\hat{M}$ and let us consider its projection onto the tangent space
$T(\Sigma)$ of $\Sigma$: $\mathcal{\ }$%
\begin{equation}
\mathcal{\mathcal{A}}^{\alpha}=e_{A}^{\alpha}\mathcal{\ \mathcal{A}}^{A}
\label{Aalfa}%
\end{equation}

With the help of (\ref{alfabeta}) and (\ref{Aalfa}) it is not difficult to
show that\ a gauge transformation of the potential, $\mathcal{\mathcal{A}}%
^{A}\rightarrow\mathcal{\ \mathcal{A}}^{A}+\hat{g}^{AB}\partial_{B}f$, induces
the gauge transformation of the projected potential $\mathcal{\mathcal{A}%
}^{\alpha}\rightarrow\mathcal{\ \mathcal{A}}^{\alpha}+g^{\alpha\beta}%
\partial_{\beta}f.$

In order to make contact with four-dimensional electrodynamics let us suppose
the existence, in addition to the projected potential $\mathcal{\mathcal{A}%
}^{\alpha}$, of an "intrinsic" electromagnetic potential $A^{\alpha}$ confined
to the spacetime and that could be measured by four-dimensional observers
living on $\Sigma$. It is then reasonable to assume a kind of "isometric"
condition for the potential holds, i.e., we assume that the intrinsic
potential $A^{\alpha}$ should be identified to $\mathcal{\mathcal{A}}^{\alpha
}$. That is
\begin{equation}
A^{\alpha}=\mathcal{\ \mathcal{A}}^{\alpha}=e_{A}^{\alpha}%
\mathcal{\ \mathcal{A}}^{A} \label{isopotential}%
\end{equation}

In analogy with four-dimensional electrodynamics the higher-dimensional
electromagnetic tensor $\mathcal{F}_{AB}$ is defined as $\mathcal{F}%
_{AB}=D_{A}\mathcal{A}_{B}-D_{B}\mathcal{A}_{A}$, while the intrinsic
four-dimensional electromagnetic tensor keeps its usual form$\ F_{\alpha\beta
}=\nabla_{\alpha}A_{\beta}-\nabla_{\beta}A_{\alpha}$. Let us now compare the
projection of $\mathcal{F}_{AB}$ relative to the submanifold $\Sigma$ with the
"intrinsic" electromagnetic tensor $F_{\alpha\beta}$. From (\ref{DBVA}) we can
calculate the covariant derivative of the higher-dimensional potential, which,
projected into the hypersurface, gives:%
\begin{equation}
e_{\beta}^{B}e_{\gamma}^{C}D_{B}\mathcal{A}_{C}=\nabla_{\beta}\mathcal{A}%
_{\gamma}-\mathcal{A}^{\bot}K_{\gamma\beta}=\nabla_{\beta}A_{\gamma
}-\mathcal{A}^{\bot}K_{\gamma\beta}%
\end{equation}
Notice that the second equality is obtained from the 'isometric' condition
(\ref{isopotential}), which allows us to substitute $\mathcal{A}_{\gamma}$ by
the intrinsic potential $A_{\gamma}$. Using the above equation and taking into
account that the extrinsic curvature tensor of $\Sigma$ is symmetric it
follows that%
\begin{equation}
\mathcal{F}_{\alpha\beta}=e_{\alpha}^{A}e_{\beta}^{B}\mathcal{F}%
_{AB}=F_{\alpha\beta} \label{Gauss}%
\end{equation}

In a geometric view of gauge theories we may regard the potential as playing
the role of a connection, while the field tensor is the associated curvature
tensor. Thus, the above condition relating the projected curvature and the
intrinsic curvature may be viewed as the analogue of the Gauss equation which
gives a relation of the Riemannian curvature tensors of the embedded manifold
and the ambient space.

At this point, let us admit that \ the electromagnetic field equations in
$\hat{M}$ are given by a higher-dimensional version of the Maxwell equations,
that is, let us assume that the dynamics of $\mathcal{\mathcal{A}}^{A}$ is
given by%
\begin{equation}
D_{A}\mathcal{F}^{AB}=g^{2}J^{B} \label{5eq}%
\end{equation}
where $g^{2}$ is the higher-dimensional coupling constant and $J^{B}$ is a
higher-dimensional electric density current that is supposed to satisfy the
continuity equation $D_{B}J^{B}=0$. The higher-dimensional field equation can
be decomposed in parallel and orthogonal parts relatively to $\Sigma$ in much
the same way as was done in (\ref{divT}). Thus, using (\ref{divT}) and
(\ref{Gauss}), the parallel projection yields the induced field equations in
terms of the intrinsic electromagnetic field $F^{\alpha\beta}$%

\begin{equation}
\nabla_{\alpha}{}F^{\alpha\beta}=g^{2}J^{B}e_{B}^{\beta}+\psi^{\alpha}\left(
K_{\alpha}^{\beta}-\delta_{\alpha}^{\beta}K\right)  -W^{\beta} \label{IE}%
\end{equation}
where $W^{\beta}=\left(  D_{\perp}T^{AB}\right)  N_{A}e_{B}^{\beta}$ . Note
that here we have used the anti-symmetry of the tensor $\mathcal{F}^{AB}$,
which allows us to write $\chi^{\alpha}=-\psi^{\alpha}$ (see (\ref{psi}) and
(\ref{qi}))

The sum of the terms on the right-hand side of the above equation, which
behaves as a vector field of $\Sigma$ with respect to the coordinates
transformations of the hypersurface, appears in (\ref{IE}) as an effective
density of current. It can be checked that this induced current has the
important property of being divergenceless. It is also worthy of mention that
the extrinsic curvature of the hypersurface contributes as a source of the
four-dimensional electromagnetic field.

Finally, the orthogonal projection of the field equation (\ref{5eq}) gives the
following constraint for the vector field $\psi^{\alpha}:$%
\begin{equation}
\nabla_{\alpha}\psi^{\alpha}=g^{2}J^{A}N_{A}%
\end{equation}

\section{Existence of embedded solutions of the induced equation}

As we have already pointed out, this new formulation of the electromagnetism
raises the fundamental question of whether all known solutions of the usual
four-dimensional Maxwell equations are contained in the induced equations. We
are going to consider this question here following an approach which bears a
close analogy with the formulation of the Campbell-Magaard embedding theorem
\cite{campbell}.

Let $A^{\alpha}\left(  x\right)  $ be a given four-dimensional potential and
$p\in\Sigma$. We want to show that there exists a solution $\mathcal{A}%
^{A}\left(  y\right)  $ of the higher-dimensional Maxwell equations
(\ref{5eq}) and an open set $U\subset\Sigma$ containing the point $p$, such
that%
\begin{equation}
\left.  e_{A}^{\alpha}\mathcal{A}^{A}\right\vert _{U}=A^{\alpha}%
\end{equation}
i.e. the projection of the 5D-solution onto the tangent space $T(\Sigma)$ of
the hypersurface is equal to intrinsic potential $A^{\alpha}\left(  x\right)
$ given.

As far as the existence of solutions is concerned there is no loss of
generality if we admit that the field equations in the higher-dimensional
space are the sourceless Maxwell equations
\begin{equation}
D_{A}\mathcal{F}^{AB}=0 \label{noJ}%
\end{equation}
\qquad

Now, if we prove that it is possible to carry out an "isometric embedding" of
any \ four-dimensional potential $\mathcal{\ }A^{\alpha}$ into $\hat{M}$, then
clearly we have demonstrated that any arbitrary of four-dimensional (physical)
current can be simulated, or rather generated, by the embedding mechanism.

Let us start the proof. Since $\mathcal{F}^{AB}$ is an anti-symmetric tensor,
it satisfies identically, i.e. independent of any field equation, the
condition:%
\begin{equation}
D_{B}D_{A}\mathcal{F}^{AB}=0 \label{Id}%
\end{equation}
Using the expression (\ref{divT}) for the anti-symmetric tensor $\mathcal{F}%
^{AB}$ we get the following identity:%
\begin{equation}
D_{B}\left[  e_{\beta}^{B}\left(  \nabla_{\alpha}F^{\alpha\beta}-j^{\beta
}\right)  +n^{B}\left(  \nabla_{\alpha}\psi^{\alpha}\right)  \right]  =0
\label{ID-foliation}%
\end{equation}
Since the equations (\ref{noJ}) and (\ref{Id}) are prescribed in an open set
of $\hat{M}$, then equation (\ref{ID-foliation}) must be also valid in
$\hat{M}$. Therefore, in (\ref{ID-foliation}) we must interpret $F^{\alpha
\beta},j^{\beta},n^{B},\nabla_{\alpha}\psi^{\alpha}$ as extensions of these
quantities which were originally defined in the hypersurface $\Sigma$. In the
sequel we shall denote these extensions with a bar. Now, an extension can be
obtained by defining a foliation of $\hat{M}$. A foliation adapted to the
embedding may be given by considering a family of functions $f^{A}\left(
x,\ell\right)  $ which depend on the coordinates of $\Sigma$ and on the extra
coordinate $\ell$, with the condition
\begin{equation}
f^{A}\left(  x,0\right)  =\phi^{A}\left(  x\right)  \label{l=0}%
\end{equation}
where $\phi^{A}\left(  x\right)  $ is the original embedding map. If
additionally we take $f^{A}\left(  x,\ell\right)  $ to be differentiable at
$p$ (at least with the same class of differentiability of $\hat{M}$), then the
set of equations%
\begin{equation}
y^{A}=f^{A}\left(  x,\ell\right)  \label{NC}%
\end{equation}
may be regarded as a legitimate transformation of coordinates in a
neighborhood of $p$; hence the set of variables $\left\{  x,\ell\right\}  $
will constitute a coordinate system in a neighborhood of $p$ $\in$ $\hat{M}$.
Clearly, each equation $\ell=const$ defines a hypersurface of $\hat{M}$. In
particular, $\ell=0$ is the equation that defines the hypersurface $\Sigma$ by
virtue of (\ref{l=0}). Naturally the mathematical formalism developed
previously is also applicable to any leaf $\left(  \ell=const\right)  $ of the
foliation, with the simple change
\begin{equation}
e_{\alpha}^{A}\left(  x\right)  \rightarrow\bar{e}_{\alpha}^{A}\left(
x,\ell\right)  =\frac{\partial f^{A}}{\partial x^{\alpha}}%
\end{equation}
Thus, in equation (\ref{ID-foliation}) the projected quantities will also
depend on the extra coordinate $\ell.$ When $\ell=0$ they reduce to the
original ones defined on the hypersurface $\Sigma$.

Now if we admit that equation (\ref{IE}) holds in a neighborhood of $p$ $\in$
$\hat{M}$, then equation (\ref{ID-foliation}) reduces to%
\begin{equation}
\bar{n}^{B}D_{B}\left(  \bar{\nabla}_{\alpha}\bar{\psi}^{\alpha}\right)
-\left(  \bar{\nabla}_{\alpha}\bar{\psi}^{\alpha}\right)  D_{B}\bar{n}^{B}=0
\label{Evol_constraint}%
\end{equation}
The above equation gives the propagation of $\nabla_{\alpha}\psi^{\alpha}$
along the normal direction. If $\nabla_{\alpha}\psi^{\alpha}=0$ in $\Sigma$,
then clearly the unique solution of (\ref{Evol_constraint}) is $\bar{\nabla
}_{\alpha}\bar{\psi}^{\alpha}=0$ in a open set of $\hat{M}.$ Therefore, as
usual in gauge theories, there is a kind of constraint equation,
$\nabla_{\alpha}\psi^{\alpha}=0$, that is automatically propagated by the real
dynamical field equations (\ref{IE}). We might say then that the field
equations $D_{A}\mathcal{F}^{AB}=0,$ defined in $\hat{M}$, is equivalent to%
\begin{align}
\bar{\nabla}_{\alpha}{}\bar{F}^{\alpha\beta}  &  =\bar{j}^{\beta}%
\qquad\text{in }\hat{M}\label{dyn}\\
\nabla_{\alpha}\psi^{\alpha}  &  =0\qquad\text{in }\Sigma\label{constraint}%
\end{align}
Now let us investigate the existence of solutions of (\ref{dyn}). With this
purpose let us rewrite the above equation in a more convenient form. We begin
by showing that $\bar{W}^{\beta}$ is the only term in equation (\ref{dyn}%
)\ that involves the second derivative of the potential along the normal
direction. In order to verify this we rewrite $\bar{W}^{\beta}$ as the
following decomposition:
\begin{equation}
\bar{W}^{\beta}=-D_{\bot}\bar{\psi}^{\beta}-\bar{F}^{\alpha\beta}%
{\LARGE a}_{\alpha}-\bar{\psi}^{\alpha}\bar{e}_{A}^{\beta}\theta_{\alpha}%
^{A}+\bar{\psi}^{\alpha}\bar{K}_{\alpha}^{\beta} \label{W}%
\end{equation}
where we define ${\LARGE a}^{\alpha}=e_{A}^{\alpha}D_{\bot}N^{A}$
(acceleration of the normal vector $\mathbf{N}$) and $\theta_{\alpha}%
^{A}=\left[  \mathbf{N},\partial_{\alpha}\right]  ^{A}$ (the Lie derivative of
$\mathbf{N}$ and the extended vector fields $\partial_{\alpha}$). It is clear
that ${\LARGE a}^{\alpha}$ and $\theta_{\alpha}^{A}$ depend on the foliation,
which is assumed to be sufficiently smooth. Consider now the term $D_{\bot
}\bar{\psi}^{\beta}$. Writing the normal vector in the coordinate basis
$\left\{  \partial_{\alpha},\partial_{\ell}\right\}  $ and using (\ref{DA}),
we can show that%

\begin{equation}
D_{\bot}\bar{\psi}^{\beta}=\mathbf{N}\left(  \bar{\psi}^{\beta}\right)
=\frac{1}{N_{A}\ell^{A}}\left(  \frac{\partial\bar{\psi}^{\beta}}{\partial
\ell}-\ell^{A}\bar{e}_{A}^{\alpha}\partial_{\alpha}\bar{\psi}^{\beta}\right)
\label{DNpsi}%
\end{equation}
where $\ell^{A}=\frac{\partial f^{A}}{\partial\ell}$. It turns out that
$\bar{\psi}^{\beta}$ already involves a normal derivative of the potential, as
we can explicitly see by rewriting it as
\begin{equation}
\bar{\psi}^{\alpha}=g^{\alpha\gamma}\left(  \nabla_{\gamma}\mathcal{A}^{\perp
}-D_{\bot}\mathcal{A}_{\gamma}+\mathcal{A}_{C}\theta_{\gamma}^{C}\right)
\end{equation}

Therefore, as we have mentioned before, $\bar{W}^{\beta}$ depends on the
second derivative of $\mathcal{A}_{\gamma}$ with respect to the coordinate
$\ell$. If we isolate this term at the left-hand side, we can put the equation
(\ref{dyn}) in the following form:%
\begin{equation}
\frac{\partial^{2}\mathcal{A}_{\beta}}{\partial\ell^{2}}=Q_{\beta}\left(
\mathcal{A}_{\alpha},\frac{\partial\mathcal{A}_{\alpha}}{\partial\ell}%
,x,\ell\right)  \label{dyncoord}%
\end{equation}
where $Q_{\beta}$ represent functions that are analytic with respect to all
the arguments. It is important to bear in mind that the normal component of
the potential $\bar{N}^{A}$ $\mathcal{A}_{A}=\mathcal{A}^{\perp}$ may be
arbitrarily chosen and should be regarded as a known analytical function.

It turns out that according to the Cauchy-Kowalewskaya theorem equation
(\ref{dyncoord}) admits analytic solutions and that these solutions are unique
as long as the initial conditions $\left.  \partial_{\ell}\mathcal{A}_{\beta
}\right\vert _{\Sigma}$ and $\left.  \mathcal{A}_{\alpha}\right\vert _{\Sigma
}$ are specified \cite{Cauchy}. Moreover, the initial functions can be any
arbitrary analytic functions. Thus, we might choose $\left.  \mathcal{A}%
_{\alpha}\right\vert _{\Sigma}=A_{\alpha}$, in order to satisfy the isometric
embedding condition (\ref{isopotential}). In addition, we might select
functions $\left.  \partial_{\ell}\mathcal{A}_{\beta}\right\vert _{\Sigma}$
that satisfy the constraint equation, whose solution always exists. Finally,
with the solution of (\ref{dyncoord}), we are able to construct a 5D-potential
$\mathcal{A}^{A}=\bar{e}_{\alpha}^{A}$ $\mathcal{A}^{\alpha}+$ $\mathcal{A}%
^{\perp}\bar{N}^{A}$ which clearly satisfies the vacuum 5D-Maxwell equations
(\ref{noJ}). This completes the proof that the set of all analytic solutions
of the four-dimensional Maxwell equations is contained in the five-dimensional
electromagnetism. This generalizes our previous study in this subject in which
we have obtained the same result investigating the solutions of the field
equations in a flat ambient space \cite{Cruz}.

\section{The effective field equations in a brane}

In this section we are going to start by admitting that matter is localized in
a brane $\Sigma$ by some confinement mechanism \cite{rubakov}. Our purpose
here is to investigate the effects of the concentration of charges and
currents on the field equations for the electromagnetic field in the brane.

The confined 4D-current can be described by means of a Dirac delta function
with support on the hypersurface. Using a foliation adapted to the embedding,
like the foliation given in (\ref{NC}), the bulk current density $J^{B}$ can
be written as:%
\begin{equation}
J^{B}=e_{\beta}^{B}j^{\beta}\frac{\delta\left(  \ell\right)  }{N_{A}\ell^{A}}
\label{source}%
\end{equation}
where $j^{\beta}$ describes the current density of the localized charges in
$\Sigma$.

It is reasonable to expect that, by virtue of the concentration of the
electric charge in $\Sigma$, the first normal derivative of the bulk potential
is discontinuous in $\Sigma$ and that the discontinuity through $\Sigma$
should depend on $j^{\beta}$. To determine the amount of this jump, let us
integrate the equation (\ref{dyn}) in some finite region of the bulk contained
in the range $-\varepsilon<\ell<\varepsilon$. Taking into account that the
`volume' element of the bulk is $\sqrt{\left\vert \bar{g}\right\vert }%
N_{A}\ell^{A}d\ell dx^{1}...dx^{n}$ (in the coordinates adapted to the
foliation) and isolating the source term on the right-hand side, we obtain
from (\ref{IE}):%
\begin{multline}
\underset{-\varepsilon<\ell<\varepsilon}{\int...\int}\left\{  \left[  \bar
{W}^{\beta}-\bar{\psi}^{\alpha}\left(  \bar{K}_{\alpha}^{\beta}-\delta
_{\alpha}^{\beta}\bar{K}\right)  \right]  +\bar{\nabla}_{\alpha}{}\bar
{F}^{\alpha\beta}\right\}  \sqrt{\left\vert \bar{g}\right\vert }N_{A}\ell
^{A}d\ell dx^{1}...dx^{n}=\\
=g^{2}\underset{\ell=0}{\int...\int}j^{\beta}\sqrt{\left\vert g\right\vert
}dx^{1}...dx^{n} \label{intEq}%
\end{multline}
\bigskip

All the terms inside the integral on the left hand side of equation
(\ref{intEq}) are bounded functions, except $\bar{W}^{\beta}$ which contains a
second derivative of the bulk potential with respect to the normal direction,
as we have seen in the previous section. Therefore $\bar{W}^{\beta}$ might
contain a singular distribution. Now taking the limit $\varepsilon
\rightarrow0$, we find, using (\ref{W}) and (\ref{DNpsi}), that the equation
(\ref{intEq}) yields
\begin{equation}
\left[  \psi^{\beta}\right]  =-g^{2}j^{\beta}\label{junction-psi}%
\end{equation}
where $\left[  \psi^{\beta}\right]  =\psi^{\beta}\left(  \ell\rightarrow
0^{+}\right)  -\psi^{\beta}\left(  \ell\rightarrow0^{-}\right)  $. If we
admit, as usual, that the brane has $Z_{2}$ reflection symmetry then it is
reasonable to assume that $\psi^{\alpha}$ is anti-symmetric with respect to
$\Sigma$. Thus it follows that%
\begin{equation}
\psi_{\left(  0^{+}\right)  }^{\beta}=-\psi_{\left(  0^{-}\right)  }^{\beta
}=-\frac{g^{2}}{2}j^{\beta}\label{psiZ2}%
\end{equation}
Now remember that the extrinsic curvature is related to the matter content in
the brane according to the equation\footnote{The positive sign is in
accordance with our definition of the extrinsic curvature tensor (\ref{K})}:%
\begin{equation}
K_{\mu\nu}^{+}=-K_{\mu\nu}^{-}=\frac{1}{2}G_{5}\left(  S_{\mu\nu}-\frac{1}%
{3}g_{\mu\nu}S\right)  \label{Z2K}%
\end{equation}
where the 4D-energy-momentum tensor $S_{\mu\nu}$ can be written as a sum of
two terms: one related to the tension of the brane and the other corresponding
to the ordinary energy distribution in the brane%
\begin{equation}
S_{\mu\nu}=-\lambda g_{\mu\nu}+\tau_{\mu\nu}\label{S-tensor}%
\end{equation}
It is interesting to notice that $\psi^{\beta}$ and $K_{\mu\nu}$ are not well
defined in the brane, since they assume different values in $\Sigma$ as we
approach the hypersurface from above $\left(  \ell>0\right)  $ or from bellow
$\left(  \ell<0\right)  $. However, the product $\psi^{\beta}K_{\mu\nu}$ is
well defined in the brane. Hence the equation (\ref{IE}) gives a well defined
equation for the electromagnetic tensor $F^{\alpha\beta}$ in $\Sigma$ when we
take the limit $\ell\rightarrow0$:%
\begin{equation}
\nabla_{\alpha}{}F^{\alpha\beta}=\frac{1}{4}g^{2}G_{5}\lambda j^{\beta}%
-\frac{1}{4}g^{2}G_{5}j^{\alpha}\tau_{\alpha}^{\beta}-W_{reg}^{\beta
}\label{indEqB}%
\end{equation}
where $W_{reg}^{\beta}$ is the regular part of $\bar{W}^{\beta}$, which is
obtained in the limit. Thus we can consider that equation (\ref{indEqB})
represents the effective equation which governs the electromagnetic field as
seen by 4D-observers. Of course (\ref{indEqB}) must fulfill an important
requirement if it is to be considered a good field equation in the brane. It
should be equivalent to the Maxwell equations in the low-energy regime in
order to be consistent with the successful description of the electromagnetic
phenomena by the 4D-Maxwell equations. As direct consequence of this
requirement, it is necessary that the unknown coupling constants must obey the
constraint $g^{2}G_{5}\lambda=4e^{2}$. It follows then that the effective
equations assume the form
\begin{equation}
\nabla_{\alpha}{}F^{\alpha\beta}=e^{2}j^{\beta}-\frac{e^{2}}{\lambda}%
j^{\alpha}\tau_{\alpha}^{\beta}-W_{reg}^{\beta}\label{indEqB2}%
\end{equation}
It is obvious that, compared to the Maxwell equations, the effective equations
have two additional terms which contribute as sources of the electromagnetic
field. Let us make some comments on the interpretation of these terms and
estimate their order of magnitude.

It is generally accepted that the the order of the brane tension $\lambda$ is
much greater than the energy scale of the ordinary matter whose distribution
is described by $\tau_{\mu\nu}$. So the second term is very small compared to
the first one in the low-energy regime. Despite of its tiny effects in the
ordinary energy scale, let us try to understand how it works. From the point
of view of 4D-observers, the origin of $\frac{e^{2}}{\lambda}j^{\alpha}%
\tau_{\alpha}^{\beta}$ might be connected to a unusual coupling between
charges and fields (including the electromagnetic field itself since
$\tau_{\mu\nu}$ also contains the energy distribution of the electromagnetic
field produces by $j^{\beta}$). In a sense, this term behaves as a kind of
medium polarization, but with uncommon features. It does not vanishes even
when there is no medium at all and the charges are just spread in the vacuum.
Another peculiarity is the fact that its effects are restricted to the region
where the source is present $\left(  j^{\beta}\neq0\right)  $. To make more
precise this interpretation, let us write the density current as $j^{\beta
}=\rho_{0}U^{\beta}$, where $\rho_{0}$ is the proper charge density and
$U^{\beta}$ corresponds to the quadrivelocity of the charges $\left(
U^{a}U_{\alpha}=-1\right)  $. Taking the inner product with $U_{\beta},$ we
can see that the term $\frac{e^{2}}{\lambda}j^{\alpha}\tau_{\alpha}^{\beta}$
produces a change in the effective value of the charge of the source in a very
similar way of a genuine polarization. Indeed, we find
\begin{equation}
\rho_{eff}=-U_{\beta}\left(  j^{\beta}-\frac{1}{\lambda}j^{\alpha}\tau
_{\alpha}^{\beta}\right)  =\rho_{0}\left(  1+\frac{u}{\lambda}\right)
\end{equation}
where $u=\tau^{\alpha\beta}U_{\alpha}U_{\beta}$ is the energy density as
measured by a co-moving inertial frame relative to the charges. If equation
(\ref{indEqB2}) does not allow solutions with negative energy distribution,
then we can conclude that due to the `polarization' the effective charge
appears to be greater than the real charge. Therefore, this
\'{}%
polarization' has an opposite sign in comparison with a common polarization of
a material medium.

The third term is a vector field $W_{reg}^{\beta}$, which might not be put in
correspondence with any known field by the 4D-observers. The reason is that it
depends on the behavior of the gauge field in the vicinity of $\Sigma$.
However, roughly speaking, it could be considered as the effect of a strange
vacuum polarization in the brane, although this vacuum polarization would have
no connection with the quantum fluctuation of the vacuum. Now let us estimate
the magnitude of $W_{reg}^{\beta}$. As we shall see it requires a more careful
examination. First, taking the divergence of the effective equation, we find
that $W^{\beta}$ satisfies the condition
\begin{equation}
\nabla_{\beta}W_{reg}^{\beta}=-\frac{e^{2}}{\lambda}\tau_{\alpha}^{\beta
}\nabla_{\beta}j^{\alpha}%
\end{equation}
It is clear that the divergence of $W^{\beta}$ has the same order of the
second term in equation (\ref{indEqB2}) and, thus can be neglected when
compared to $e^{2}j^{\alpha}$. However, the divergenceless part of $W^{\beta}$
cannot be estimated by 4D-observers. This is because the induced equations do
not constitute a complete set of equations for the 4D-fields. They can be
solved only if they are considered as part of the five-dimensional field
equations. However, some analyses regarding specific geometric configurations
show that the term cannot be discarded, and, as consequence the 4D-Green
function cannot be recovered in the brane at low-energy regime
\cite{pomarol,bajc}.

This negative result gives a clear indication that massless vectors fields
cannot be localized in the hypersurface by means of gravity in this scenario.
However, as we have already mentioned there are some alternative models which
propose other different ways to guarantee the localization of massless gauge
field. Here we want to concentrate our discussion on the DGP model.

According to DGP model, the interaction between bulk gauge fields and
localized matter in the brane induces corrections in the original Lagrangean
due to quantum effects. It is argued that these additional terms represent
interactions which are already four-dimensional. For example, in the case of
electromagnetism, the total Lagrangean gains an extra term which is exactly
$\frac{1}{e^{2}}F_{\alpha\beta}F^{\alpha\beta}$, the usual 4D-electromagnetic
Lagrangean, restricted to the world volume of the brane \cite{DGP2}. Of
course, the presence of this extra term will modify the field equations. Let
us see that this alteration is easily implemented within our formalism.
Indeed, as we have mentioned $F_{\alpha\beta}F^{\alpha\beta}$ is already
confined to the brane, as the original source $j^{\beta}$. Thus,
$F_{\alpha\beta}F^{\alpha\beta}$ can be readily incorporated into the field
equations if we take it as part of the bulk current $J^{B}$. Thus the new bulk
current now becomes:
\begin{equation}
J^{B}=e_{\beta}^{B}\left(  j^{\beta}-\frac{1}{e^{2}}\nabla_{\alpha}%
F^{\alpha\beta}\right)  \frac{\delta\left(  \ell\right)  }{N_{A}\ell^{A}%
}\label{Source-DGP}%
\end{equation}
Following the previous procedure, it can be readily seen that the
discontinuity of the $\bar{\psi}^{\beta}$ through $\Sigma$ is now given by:%
\begin{equation}
\left[  \psi^{\beta}\right]  =-g^{2}\left(  j^{\beta}-\frac{1}{e^{2}}%
\nabla_{\alpha}F^{\alpha\beta}\right)  \label{Junction-psi-DGP}%
\end{equation}

Taking the equation (\ref{IE}) in the limit $\ell\rightarrow0$ and admitting
the reflection symmetry again we obtain the effective field equation in the
brane in DGP scenario:
\begin{equation}
\left\{  \left(  1+\frac{1}{4}\frac{g^{2}\lambda G_{5}}{e^{2}}\right)
\delta_{\gamma}^{\beta}-\frac{1}{4}\frac{g^{2}G_{5}}{e^{2}}\tau_{\gamma
}^{\beta}\right\}  \nabla_{\alpha}F^{\alpha\gamma}=\frac{1}{4}g^{2}\lambda
G_{5}j^{\beta}-\frac{1}{4}g^{2}G_{5}j^{\alpha}\tau_{\alpha}^{\beta}%
-W_{reg}^{\beta} \label{EqDGP}%
\end{equation}

As in the previous case, in order that equation (\ref{EqDGP}) can give an
appropriate low-energy limit, it is necessary that the coupling constant be
constraint according to this formula%
\begin{equation}
g^{2}\lambda G_{5}=4\sigma e^{2}%
\end{equation}
where $\sigma$ is some dimensionless constant (in our units). Substituting
this constraint into the (\ref{EqDGP}), the effective field equations assume
the following form:%
\begin{equation}
\left[  \delta_{\gamma}^{\beta}-\frac{\sigma}{\left(  1+\sigma\right)
\lambda}\tau_{\gamma}^{\beta}\right]  \nabla_{\alpha}F^{\alpha\gamma}%
=\frac{\sigma e^{2}}{\left(  1+\sigma\right)  }j^{\beta}-\frac{\sigma e^{2}%
}{\left(  1+\sigma\right)  \lambda}j^{\alpha}\tau_{\alpha}^{\beta}-\frac
{1}{\left(  1+\sigma\right)  }W_{reg}^{\beta}\label{EqDGP2}%
\end{equation}
There are two terms that depend on the brane tension $\lambda$ directly. As we
have pointed out before, they are very small compared to first term
($e^{2}j^{\beta})$ at the ordinary scale of energy. So let us focus our
analysis on $W_{reg}^{\beta}$. Our first remark is that now, in the DGP
scenario, the vector field is constrained. Indeed, comparing equation
(\ref{Junction-psi-DGP}) and (\ref{EqDGP2}), we find that:
\begin{equation}
W_{reg}^{\beta}=-e^{2}j^{\beta}+\left(  1+\sigma\right)  \left(
\delta_{\gamma}^{\beta}-\frac{1}{2\lambda}\tau_{\gamma}^{\beta}\right)
\frac{e^{2}}{g^{2}}\left[  \psi^{\gamma}\right]
\end{equation}
This equation shows that there is now two different characteristic parameters:
the brane tension $\lambda$ and the electromagnetic coupling constant $g^{2}$.
We have already investigated the limit of the effective equations in
comparison with the energy scale of $\lambda.$ Now let us consider the other
parameter $g^{2}$. If we admit that the lenght scale $\left(  G_{5}%
\lambda\right)  ^{-1}$ is much smaller than $r_{c}\equiv g^{2}/e^{2}$, which
is the most interesting case, then, in the low-energy limit (compared to the
brane tension), the above equation reduces to%
\begin{equation}
W_{reg}^{\beta}=-e^{2}j^{\beta}+\left(  1+\sigma\right)  \frac{e^{2}}{g^{2}%
}\left[  \psi^{\beta}\right]  \label{constraintDGP}%
\end{equation}

By the definition of $W^{\beta}$ and $\psi^{\beta}$, we can conclude from
(\ref{constraintDGP}), that the solution (the bulk potential) will depend on
the parameter $g^{2}$, i.e., $\mathcal{A}^{\beta}\left(  x,\ell,g^{2}\right)
$. Now if the dependence on $g^{2}$ is such that $\left[  \psi^{\beta}\right]
\sim\mathcal{O}\left[  \left(  g^{2}\right)  ^{-n}\right]  $, where $n>-1$, or
even logarithmic, then for large values of $g^{2}$, we have approximately
$W_{reg}^{\beta}=-e^{2}j^{\beta}\left(  x\right)  $, (remember that $j^{\beta
}\left(  x\right)  $ does not depend on $g^{2}$). It follows then, from
(\ref{EqDGP2}), that Maxwell equations will be recovered in the brane in the
range $\left(  G_{5}\lambda\right)  ^{-1}<<r$ $<<$ $r_{c}$.

However, for large distance compared to $r_{c}$, it is reasonable to think
that  the term $\frac{1}{r_{c}}\left[  \psi^{\beta}\right]  $ becomes more
relevant, and hence the electromagnetic field in the brane in that domain will
deviate significantly from its usual behavior in four-dimensional spacetime.
In other words, the five-dimensional character of the fundamental
electromagnetic field will be appreciable at large distance compared to the
scale established by the parameter $r_{c}$. It is well known that long range
effects are connected to low frequency modes. Therefore, the deviation of the
four-dimensional behavior is interpreted as a consequence of the leakage of
infrared modes of the field to the extra dimension. This effect is known as
infrared transparency and it is a characteristic feature of the DGP model.

\section{Final remarks}

Originally the DGP model was conceived to deal with the issue of localization
of gravitons in brane worlds. According to this model, the Newtonian
gravitational potential can be recovered in a brane even if the brane is
embedded in a bulk with flat extra dimension. The key element, as the authors
realized, is the fact that matter localized in the brane interacting with bulk
gauge fields gives rise to a kinetic term for the gauge fields which is
constrained to the world volume of the brane. Soon afterwards, this scheme was
applied to other gauge interactions as a generic mechanism of localization of
massless gauge fields. A curious feature of the DGP model is that deviation of
the four-dimensional behavior of the localized field will be manifest at
ultra-large distance.

In this paper we have considered an extended version of DGP braneworld
scenario and investigated the question of how the electromagnetism is
described in the brane. Using a covariant embedding formalism, we have found
the effective field equations that dictated the behavior of the
electromagnetism field as seen from 4D-observers. The adopted formulation here
seems to have some advantages over some other approaches: First, it allows us
to treat the issue with great generality since we do not put restrictions on
the geometry of neither the bulk nor the brane. Second, based on a covariant
formulation, the interpretation of the terms that appear in the effective
field equations are independent of coordinates choices.

In order to keep the maximum generality we have also considered branes whose
tension is not necessarily null. Hence, the effective field equations have two
parameter which establish two different length scales. Finally, we have
determined the condition under which the effective field equations can recover
Maxwell equations.

\section{Acknowledgement}

The authors would like to thank CNPq-FAPESQ (PRONEX) for financial support.

\end{document}